\documentclass[conference]{IEEEtran}
\IEEEoverridecommandlockouts
\usepackage{cite}
\usepackage{amsmath,amssymb,amsfonts}
\usepackage{algorithmic}
\usepackage{graphicx}
\usepackage{textcomp}
\usepackage[caption=false]{subfig}
\usepackage{makecell}

\usepackage{xcolor}
\def\BibTeX{{\rm B\kern-.05em{\sc i\kern-.025em b}\kern-.08em
    T\kern-.1667em\lower.7ex\hbox{E}\kern-.125emX}}

\graphicspath{ {./Images/} } 

\def\x{{\mathbf x}}
\def\z{{\mathbf z}}
\def\c{{\mathbf c}}
\def\xhat{{\hat{\mathbf x}}}
\def\B{{\mathbf B}}
\def\S{{\mathbf S}}
\def\L{{\mathbf L}}
\def\W{{\mathbf W}} 
\def\D{{\mathbf D}} 
\def\U{{\mathbf U}}
\def\p{{\mathbf p}}
\def\q{{\mathbf q}}
\def\s{{\mathbf s}}
\def\P{{\mathbf P}}

\def\calG{{\cal G}}
\def\calV{{\cal V}}
\def\calE{{\cal E}}

\begin{document}

\title{Adaptive Online Learning of Separable Path Graph Transforms for Intra-prediction\\

\thanks{This work is funded in part by a gift from Tencent America LLC.}
}

\author{
    \IEEEauthorblockN{Wen-Yang Lu, Eduardo Pavez, Antonio Ortega}
    \IEEEauthorblockA{University of Southern California\\
      Department of ECE\\
      Los Angeles, CA, USA}
    \and
    \IEEEauthorblockN{Xin Zhao, Shan Liu}
    \IEEEauthorblockA{Tencent Media Lab\\ 
                      Palo Alto, CA, USA\\}
}
\maketitle

\begin{abstract}
Current video coding standards, including H.264/AVC, HEVC, and VVC, employ discrete cosine transform (DCT), discrete sine transform (DST), and secondary Karhunen–Loève transforms (KLTs) to decorrelate the intra-prediction residuals.
However, the efficiency of these transforms in decorrelation can be limited when the signal has a non-smooth and non-periodic structure, such as those occurring in textures with intricate patterns. 
This paper introduces a novel adaptive separable path graph-based transform (GBT) that can provide better decorrelation than the DCT for intra-predicted texture data. 
The proposed GBT is learned in an online scenario with sequential \textit{K}-means clustering, which groups similar blocks during encoding and decoding to adaptively learn the GBT for the current block from previously reconstructed areas with similar characteristics. 
A signaling overhead is added to the bitstream of each coding block to indicate the usage of the proposed graph-based transform. 
We assess the performance of this method combined with H.264/AVC intra-coding tools  
and demonstrate that it can significantly outperform H.264/AVC DCT for intra-predicted texture data.
\end{abstract}


\section{Introduction}
In video coding, exploiting spatial redundancy to enhance coding efficiency is a fundamental principle, primarily achieved through techniques like intra-prediction and signal transforms. 
Prominent video coding standards, including H.264/AVC \cite{wiegand2003overview}, HEVC \cite{sullivan2012overview}, and VVC \cite{bross2021overview}, have conventionally relied on well-established transforms such as the discrete cosine transform (DCT) and the discrete sine transform (DST) \cite{strang1999discrete,zhao2019vvc} to decorrelate intra-prediction residuals effectively.
However, transform kernels are all fixed and hardcoded in these standards; thus, these transforms have limited effectiveness for non-smooth and non-periodic structures, such as images with intricate textures. 


To fully exploit the data correlations, signal-dependent transforms (SDT) have been developed to customize the transform kernel according to the unique properties of each block. 
The Karhunen-Loève Transform (KLT), the optimal SDT for energy compaction and signal decorrelation, is extensively utilized in signal processing and statistical analysis.
Intra-frame coding with the KLT was investigated in \cite{ye2008improved, takamura2013intra, liu2018scene}, where the KLT kernels are trained offline and adapt to each of the directional modes to avoid the high computational costs associated with learning during the encoding process. 
Nevertheless, for video codecs with many directional prediction modes, such as HEVC and VVC, using numerous transform kernels could pose challenges for memory and hardware design.
An online-learned KLT approach using template matching was proposed in \cite{lan2011exploiting}, where only one bit of overhead is needed to signal whether the SDT is used for each block. 
Every block to be encoded is represented by a template, namely, a set of top and left neighboring previously coded pixels (see Fig.~\ref{fig:template}). 
Then, the $M$ most similar blocks from the previously reconstructed area are searched by comparing the templates of candidate blocks with that of the current block. A KLT kernel associated with the current block is obtained by eigendecomposition of the empirical covariance matrix calculated from these $M$ blocks.
Compared with the offline trained KLT adapted to the directional characteristics of prediction modes, this online scenario ensures the adaptability to the data itself and better exploits the data correlations.
However, learning the KLT requires a reliable covariance estimate, which can only be achieved with a sufficiently large number of observations.
In addition, finding the $M$ most similar blocks requires a significant amount of template comparisons (typically several hundred) for each block.

Graph Signal Processing (GSP) \cite{shuman2013emerging, ortega2018graph} considers problems widely studied in conventional signal processing and solves them in the context of graphs. 
Notions of frequency and graph Fourier transform, analogous to the conventional Fourier transform, have been developed to analyze the signal's characteristics, such as smoothness, associated with a given graph. 
Several algorithms for learning graphs as an approximation to inverse covariance matrices (precision matrices) have been developed  
\cite{egilmez2017graph, pavez2018learning, lu2018learning, mateos2019connecting, dong2019learning}.
Since the eigenvectors of the covariance matrix form the KLT and 
the graph Laplacian of the learned graph approximates the inverse covariance matrix,  
the eigenvectors of the graph Laplacian, i.e., the graph Fourier transform (GFT), can be viewed as approximations to the KLT.
It is well-known that for path graphs with unit edge weights (with possibly some self loops) the KLT/GFT are the ADST and DCT, respectively \cite{strang1999discrete}.
In \cite{lu2018learning,pavez2018learning}, it was shown that tree graphs can be learned from data in a closed form. Since path graphs can be used to construct 2D separable transforms (as is done with the DCT and DST), GFTs for path graphs can be viewed as generalizations of traditional transforms used in video coding.
In addition, GFTs of path graphs learned from covariance matrices, i.e., approximations of the KLT, can often outperform the KLT \cite{egilmez2020graph}. This is because path graph learning converges at faster rates and requires learning fewer parameters \cite{pavez2022laplacian}. 
Thus, constraining the graph to be a path graph acts as some form of regularization. 
This is particularly advantageous when the amount of data available for training is small.
In all of the previous work \cite{pavez2018learning, mateos2019connecting, dong2019learning, lu2018learning, egilmez2020graph}, graph-based transforms (GBT) are learned offline (i.e., based on training data) and are made available at both the encoder and the decoder. 

This paper proposes an online scenario to learn block adaptive separable path GBTs. Our method addresses two of the main drawbacks of learning KLT online by template matching in \cite{lan2011exploiting}: 
\paragraph*{1) Complexity} 
Learning the KLT kernels online by template matching requires finding the $M$ nearest templates, which involves computing distances to many candidates (several hundreds in general) for each block. 
In contrast, we propose a sequential $K$-means clustering method, which only requires $K$ comparisons (one per cluster) per block, where $K$ can be chosen to be small, e.g., 8 or 16.
Furthermore, in \cite{lan2011exploiting}  the covariance matrix for each $n \times n$ block has $O(n^4)$ parameters, which requires $O(Mn^4)$ time when estimated from the $M$ nearest templates. 
Instead, in our approach, the precision matrices (inverse covariance matrices) for row and column transforms have $n-1$ parameters, calculated as an average sum of squared differences in $O(n^2)$ time for each block.
\paragraph*{2) Low accuracy due to the lack of data} 
Accurate estimation of $n^2 \times n^2$ covariance matrix requires at least $M \geq n^2$, but in practice, the number of nearest templates is $M<n^2$, which results in a low-rank covariance matrix so that only a few basis vectors of the KLT can be learned from it. 
Instead of estimating the full covariance matrix, our proposed method only estimates $2(n-1)$ parameters for the vertical and horizontal path graphs. Thus, fewer data samples are required to learn the proposed GBT.

\section{Preliminaries}
\subsection{Combinatorial Graph Laplacian (CGL)}
Consider a weighted undirected graph $\calG = (\calV, \calE, \W)$ with a set of nodes $\calV$ representing pixels in an image block and a set of edges $\calE$ describing the pairwise pixel correlation.
Denote the entry $w_{i,j}$ in $\W$ as the weight for an edge $(i,j) \in \calE$  and $w_{i,j}=0$  if $(i,j) \notin \calE$. 
The combinatorial graph Laplacian (CGL) matrix is defined as 
\begin{equation}
    \label{eq_L}
    \L:=\D-\W,  
\end{equation}
where $\D$ is the diagonal degree matrix with $d_{i,i} = \Sigma^n_{j=1} w_{i,j}$.

\subsection{Separable Path Graph Transform}
Given a CGL matrix $\L$ and its eigendecomposition $\L=\U\mathbf{\Lambda}\U^\top$, the GBT of a signal vector $\x$ is obtained by $\xhat:=\U^\top\x$. A path graph is a tree with edges defined as $\calE_{\text{path}}=\{(i,j):\  |i-j|=1,\hspace{0.5em}\forall i,j \in \calV\}$, which can be drawn so that all of its vertices and edges lie on a single straight line.
Let the image block size be $n \times n$, and given two GBT operators $\U_\text{vert}$ and $\U_\text{horiz}$ associated with vertical path graph $\L_\text{vert}$ and horizontal path graph $\L_\text{horiz}$, respectively. The separable path graph transform of an image block denoted by an $n \times n$ square matrix $\B$ is obtained by
\begin{equation}
    \label{eq_separableGBT}
    \hat{\B}:=\U_\text{vert}^\top\B\U_\text{horiz},
\end{equation}
where $\U_\text{vert}$ is applied to column vectors of $\B$, and $\U_\text{horiz}$ is applied to row vectors of $\B$.

\subsection{Graph Weight Estimation}
Given the edges $\calE$ and the empirical covariance matrix $\S = \Sigma^N_{i=1}\x_i\x^\top_i/N$ from $N$ data samples $\x_i$, the goal is to estimate the CGL as formulated in \cite{egilmez2017graph}:
\begin{equation}
    \label{eq_CGLProb}
    \min _{\mathbf{L} \in \mathbb{L}(\calE)} \quad-\log \operatorname{det}\left(\mathbf{L}+\mathbf{1 1}^T / n\right)+\operatorname{trace}(\mathbf{L K}),
\end{equation}
where $\mathbf{K}=\mathbf{S}+\alpha\left(\mathbf{I}-\mathbf{1 1}^T\right)$, and $\mathbb{L}(\calE)$ is the set of CGLs with connectivity constraints defined by $\calE$ \cite{lu2018learning}.
The problem \eqref{eq_CGLProb} is a maximum
a posteriori (MAP) parameter estimation of a Gaussian Markov Random Field (GMRF) $\mathbf{x} \sim \mathcal{N}(\mathbf{0}, \boldsymbol{\Sigma}=\L^{\dagger})$.
The GBT can optimally decorrelate the signal because the covariance of the transformed signal can be shown to be a diagonal matrix. That is, $\operatorname{Cov}[\U^\top\x]=\mathbf{\Lambda}^{\dagger}$.

It was proved in \cite{lu2018learning} that if $\calE$ corresponds to a tree topology, the edge weights providing the optimal solution to \eqref{eq_CGLProb} are:
\begin{equation}
    \label{eq_closedFormW}
    w_{u,v} =\left[\underbrace{\frac{1}{N} \sum_{i=1}^N\left(\mathbf{x}_i(u)-\mathbf{x}_i(v)\right)^2}_{:=\delta_{u,v}}+2 \alpha\right]^{-1}, \quad \forall (u,v) \in \mathcal{E},
\end{equation}
where $\alpha>0$ should be set to a small positive number to avoid infinity, and $\delta_{u,v}$ is the mean-square difference (MSD) between the $u$-th and the $v$-th elements over all samples $\x_i$. The CGL matrix is then obtained by \eqref{eq_L}. 

\subsection{Sequential K-means Clustering}
Sequential $K$-means \cite{king2012online} is a variant of the traditional $K$-means clustering algorithm designed to handle large datasets more efficiently. Instead of the batch-oriented approach of standard $K$-means, sequential $K$-means processes the data incrementally, allowing for iterative updates to the cluster centroids as new data samples arrive. Given a new data sample $\x^{(t)}$ available at iteration $t$, and with the $k$-th cluster being closest to the new sample, the centroid of the $k$-th cluster is updated by
\begin{equation}
    \label{eq_updateCentroid}
    \c^{(t)}_k=\c^{(t-1)}_k+\rho(\x^{(t)}-\c^{(t-1)}_k),
\end{equation}
where $\rho$ is the learning rate (or forgetting parameter) that controls how fast centroids adapt to new data. A common choice is $\rho=0.1$, which we choose as our default setting.

\section{Proposed Method}
Our proposed separable path GBT is combined with the intra-prediction of H.264/AVC standard and is utilized as an alternative transform to DCT. 
Every image block selects the transform having better coding efficiency using rate-distortion optimization.
Overhead is sent for each block to signal which transform is adopted.
Next, we introduce our proposed online learning of graph edge weights. 

\subsection{Incremental Update of Graph Edge Weights}
In our approach, we obtain separable 2D block transforms by online learning 
vertical and horizontal path graphs.
Because path graphs are trees, their graph edge weights can be learned using  \eqref{eq_closedFormW}.
Given a collection of $M$ image blocks of size $n \times n$, we have $N=Mn$ samples (columns) for the vertical path graph and $N=Mn$ samples (rows) for the horizontal path graph. 
Let $\p_i$ be the vectors of image columns, and $\q_i$ be the vectors of image rows. The edge weights for the vertical and horizontal path graphs are learned by replacing $\x_i$ in \eqref{eq_closedFormW} with $\p_i$ and $\q_i$, respectively. 
Given a new image block, computing the new graph edge weights using \eqref{eq_closedFormW} requires updating the vertical and horizontal MSDs as:   
\begin{equation}
    \label{eq_updateMSD_vert}
    \scalebox{1.1}{$
        \delta_{u,v}^{\text{(vert)}(t)}=\frac{nM^{(t-1)}\delta_{u,v}^{\text{(vert)}(t-1)} +\sum\limits_{i=1}^{n}\left(\p_i(u)-\p_i(v)\right)^2}{n(M^{(t-1)}+1)},
    $}
\end{equation}
\begin{equation}
\label{eq_updateMSD_horiz}
    \scalebox{1.1}{$
        \delta_{u,v}^{\text{(horiz)}(t)}=\frac{nM^{(t-1)}\delta_{u,v}^{\text{(horiz)}(t-1)} +\sum\limits_{i=1}^{n}\left(\q_i(u)-\q_i(v)\right)^2}{n(M^{(t-1)}+1)}.
    $}
\end{equation}



\subsection{Adaptive Online Learning}
\begin{figure}
    \centering
    \includegraphics[width=0.4\linewidth]{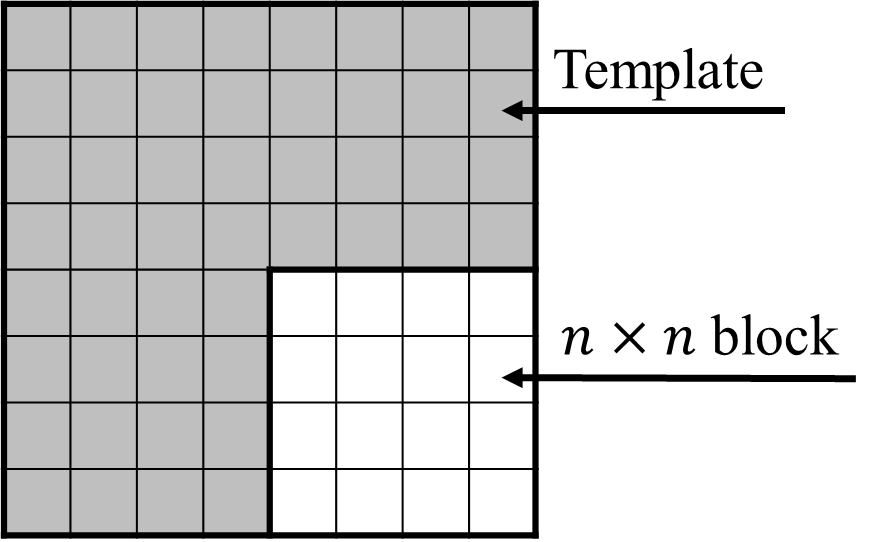}
    \caption{Illustration of the template adjacent to the image coding block.}
    \label{fig:template}
\end{figure}

To collect samples with similar pixel correlation within an image block, the sequential $K$-means clustering groups similar blocks together according to their templates, which are previously reconstructed pixels adjacent to the image block. 
An example of a template is illustrated is Fig.~\ref{fig:template}. In our paper we choose the template's width to be twice the width of the image block (other sizes would be possible). 
The vectorization of a template, $\z$,  has $3n^2$ entries in total, excluding the pixels within the image block.
Templates have the benefit that they only depend on previously coded pixels and are not involved with the current block so they can be reproduced at both the encoder and decoder without overhead.   

Each cluster learns its own path graphs and GBT. Therefore, the following information is stored and continually updated in each cluster during the learning process:
\begin{itemize}
    \item $M$: Number of block samples
    \item $\c$: Centroid of templates
    \item $\delta_{u,v}^{\text{(vert)}},\hspace{0.5em} \forall (u,v) \in \calE_\text{path}$: MSD for vertical path graph 
    \item $\delta_{u,v}^{\text{(horiz)}},\hspace{0.5em} \forall (u,v) \in \calE_\text{path}$: MSD for horizontal path graph 
    
\end{itemize}
As for the initialization, starting from the second block row and second block column of the whole image, the templates associated with the first $K$ blocks are used as the initial centroids of the $K$ clusters.
For an image block to be encoded, its associated template is built and compared to the centroids of $K$ clusters to find the closest one. 
The MSDs $\delta_{u,v}^{\text{(vert)}}$ and $\delta_{u,v}^{\text{(horiz)}}$ of the closest cluster are used to derive the separable path GBT. 
After the encoding of the current block is finished, 
the cluster is updated with:
\begin{itemize}
    \item $M^{(t)}=M^{(t-1)}+1$ 
    \item $\c^{(t)}$ is updated by \eqref{eq_updateCentroid} where $\x$ is replaced by $\z$
    \item $\delta_{u,v}^{\text{(vert)}(t)}$ is updated by \eqref{eq_updateMSD_vert}
    \item $\delta_{u,v}^{\text{(horiz)}(t)}$ is updated by \eqref{eq_updateMSD_horiz}
\end{itemize}
If the GBT is unavailable due to a lack of samples in the cluster, the DCT is used instead. Regardless of whether the GBT or DCT is used, the current reconstructed block and its template are used to update the cluster.

\subsection{Comparing with the KLT-based SDT} 
While the KLT offers optimal data decorrelation, it is computationally demanding due to the need to estimate the covariance matrix for each block.
In contrast, the separable path GBT can save a lot of computational costs because graph edge weights are learned in closed form and updated online.  

Specifically, for each $n \times n$ block, learning the KLT kernel requires several hundreds of comparisons with previously reconstructed data to identify the $M$ nearest templates and $O(Mn^4)$ time to compute the covariance matrix from the $M$ templates. 
In contrast, our proposed approach adopts the sequential $K$-means clustering, reducing the number of comparisons to only $K$ for each block, with $K$ typically set at 8 or 16. 
The graph edges weights are then learned from a closed-form \eqref{eq_closedFormW}, where the MSD are updated by \eqref{eq_updateMSD_vert} and \eqref{eq_updateMSD_horiz} with $O(n^2)$ time for each coding block. 
Moreover, when examining the convergence rate of basis learning, the path GBT tends to converge faster, which will be observed in the next section. 
This is attributed to the fact that the GBT involves learning a tree-structured graph Laplacian with few coefficients (specifically, $2(n-1)$ edge weights for the vertical and horizontal path graphs), in contrast to the full matrix $\S$ where $n^2(n^2+1)/2$ coefficients need to be learned \cite{pavez2022laplacian}.

\section{Experimental Results}

\begin{figure}[t!]
    \vspace{-5mm}
    \centering
    \subfloat[\label{sfig:evenPCorr}]
        {\includegraphics[width=0.5\linewidth]{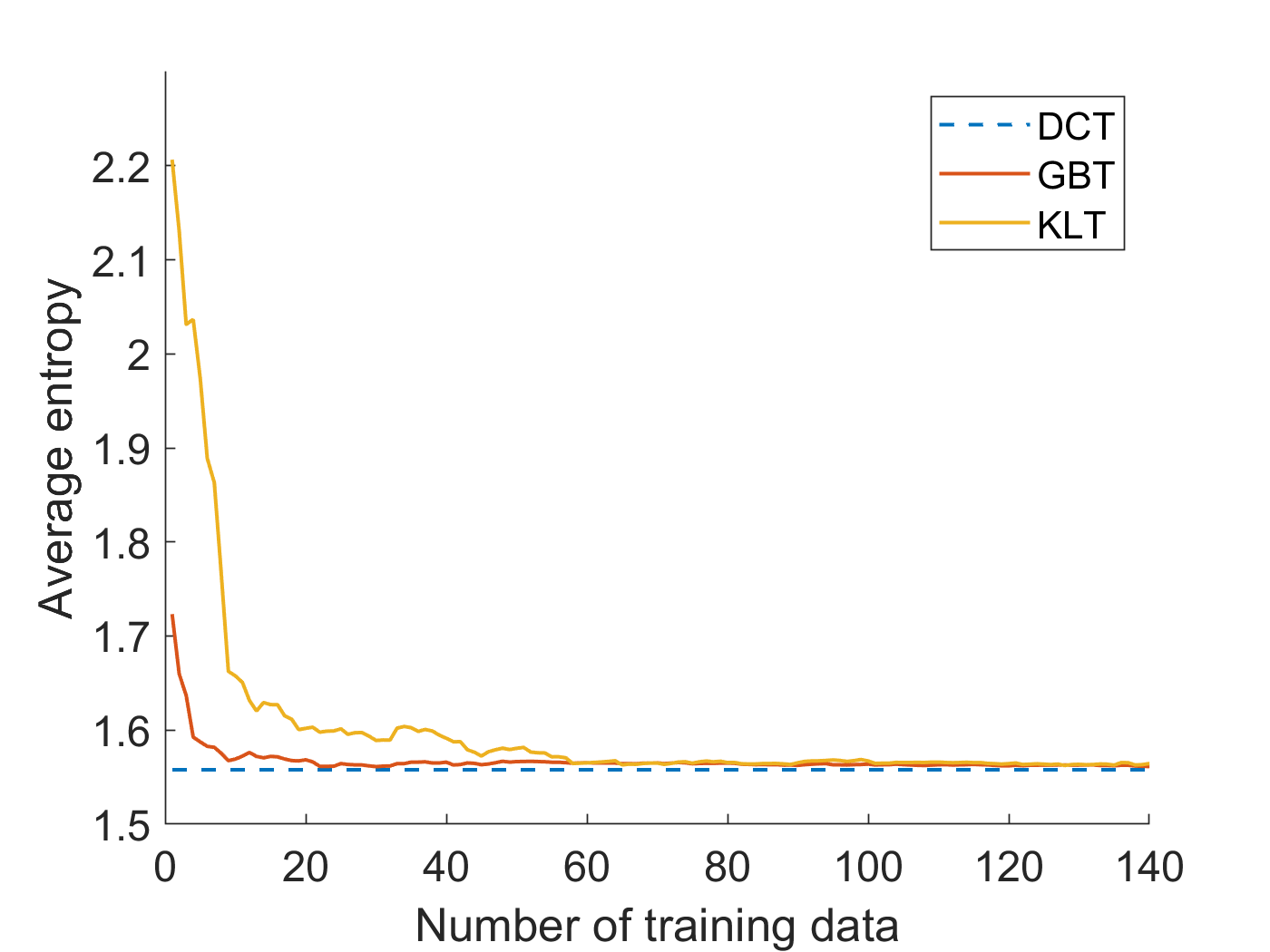}}  
    \subfloat[\label{sfig:unevenPCorr}]
        {\includegraphics[width=0.5\linewidth]{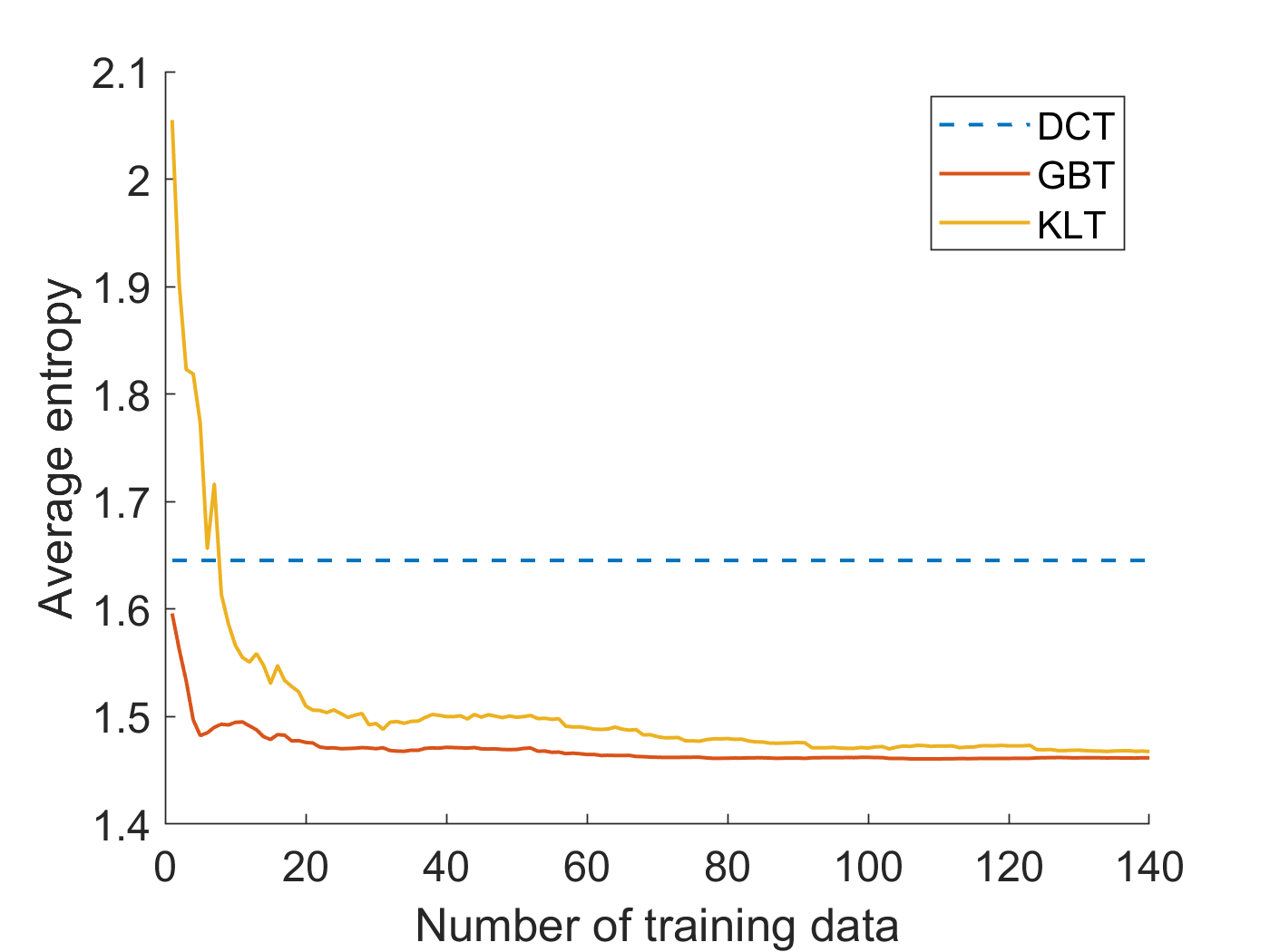}}
    \caption{Average power spectral entropy of DCT, GBT, and KLT over 1000 test samples w.r.t different number of training samples. The data models are GMRF with (a) same partial correlation; (b) different partial correlation.}
    \label{fig:GMRF}
\end{figure}

\subsection{Synthetic GMRF Signal}
This simulation aims to evaluate the effectiveness of data decorrelation and the convergence speed of the proposed path GBT. 
We simulate a GMRF model $\mathbf{x} \sim \mathcal{N}(\mathbf{0},\P^\dagger)$ and repeatedly collect samples from this model, where $\mathbf{x} \in \mathbb{R}^8$  and $\P$ is a precision matrix equal to the path graph Laplacian (eq. (15) in \cite{egilmez2020graph}), which is equivalent to the DCT \cite{egilmez2020graph}.
This implies that the data is characterized by the same partial correlation over different points.
To directly compare the GBT and KLT, the non-separable GBT is considered here with an assumed 8-point path graph topology, where the edge weights are learned from the samples via \eqref{eq_closedFormW}.
The efficiency of the DCT, KLT, and GBT are measured in terms of the power spectral entropy. Given a signal transform $\xhat$ of size $n \times 1$, its normalized power spectrum is defined as 
$\s(i)={|\xhat(i)|^2}/{\Sigma_{j=1}^n|\xhat(j)|^2}$.
The power spectral entropy (PSE) of a signal transform is obtained by 
\begin{equation}
    E=-\sum_{i=1}^n{\s(i)}\ln{\s(i)}.
\end{equation}
Lower PSE means that the signal has a more concentrated power spectrum, which is beneficial for compression.
The PSE is calculated as an average of 1000 test samples. 
Fig.~\ref{fig:GMRF}\subref{sfig:evenPCorr} shows the comparison of the average PSE between the DCT, GBT, and KLT when learning from different amounts of training data. Because the DCT is a signal-independent transform, the number of training data does not affect its performance. The GBT converges
faster than KLT, showing its advantage when there are insufficient samples to learn the transform. The GBT and KLT converge to the same performance as the DCT, showing that DCT is a special case of GBT with an undirected path graph topology.

Next, we consider a precision matrix 
with partial correlations set at 7 equidistant points, including the endpoints between $[0.1,1]$.
This implies that some points are more partially correlated than others, which could be the practical case for intricate patterns in texture images. Fig.~\ref{fig:GMRF}\subref{sfig:unevenPCorr} compares the average PSE between the DCT, GBT, and KLT. Unlike the previous simulation, it can be observed that, at convergence, both the KLT and GBT reach a lower PSE (i.e. better coding efficiency) than the DCT. This suggests that the efficiency of DCT can be limited by complex data structures.

\subsection{Intra-prediction on Real-World Texture Images}


\begin{figure}[t!]
    \vspace{-5mm}
    \centering
    \subfloat[\label{sfig:PSNR}]
      {\includegraphics[width=0.5\linewidth]{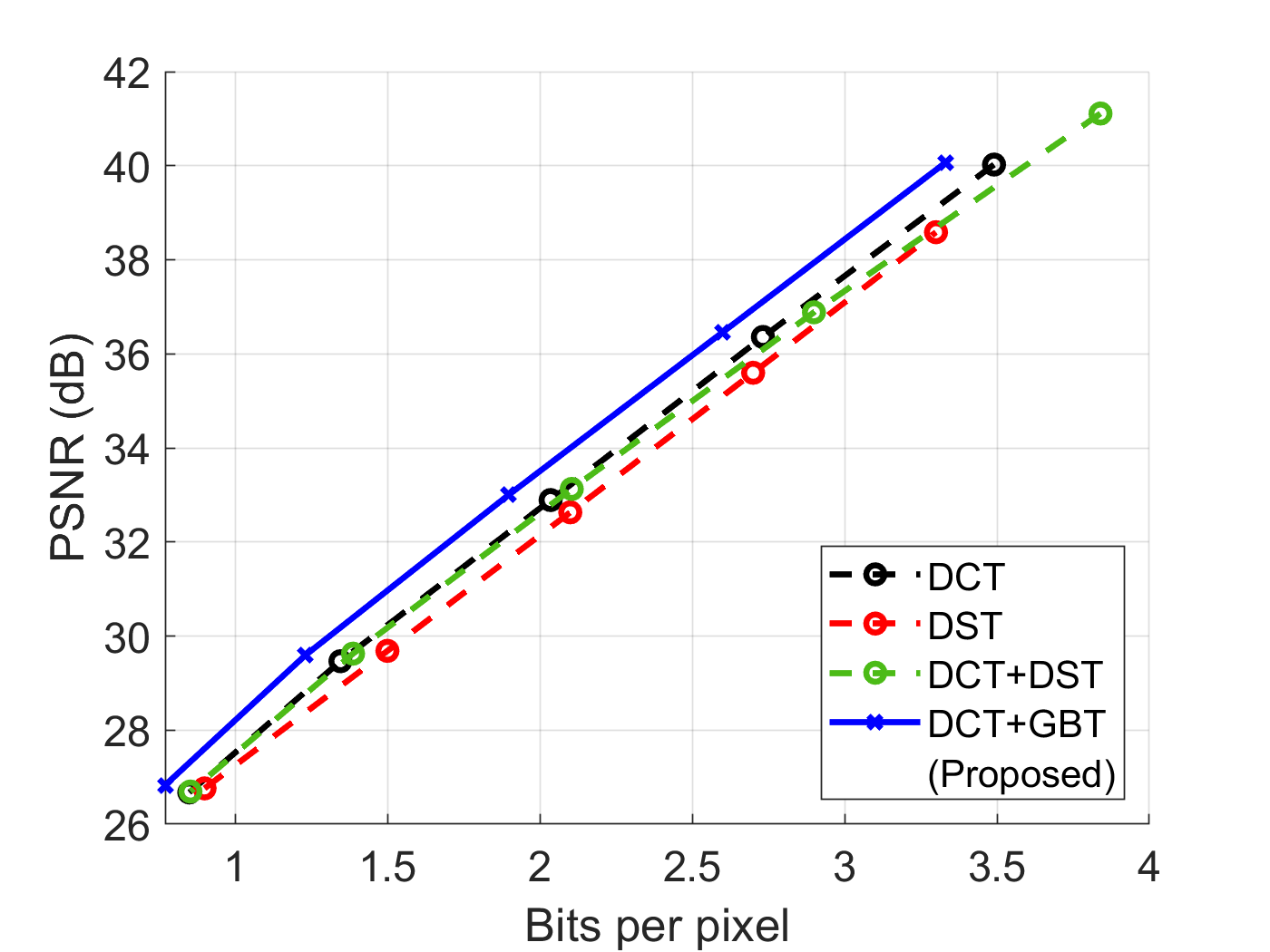}}  
    \subfloat[\label{sfig:SSIM}]
       {\includegraphics[width=0.5\linewidth]{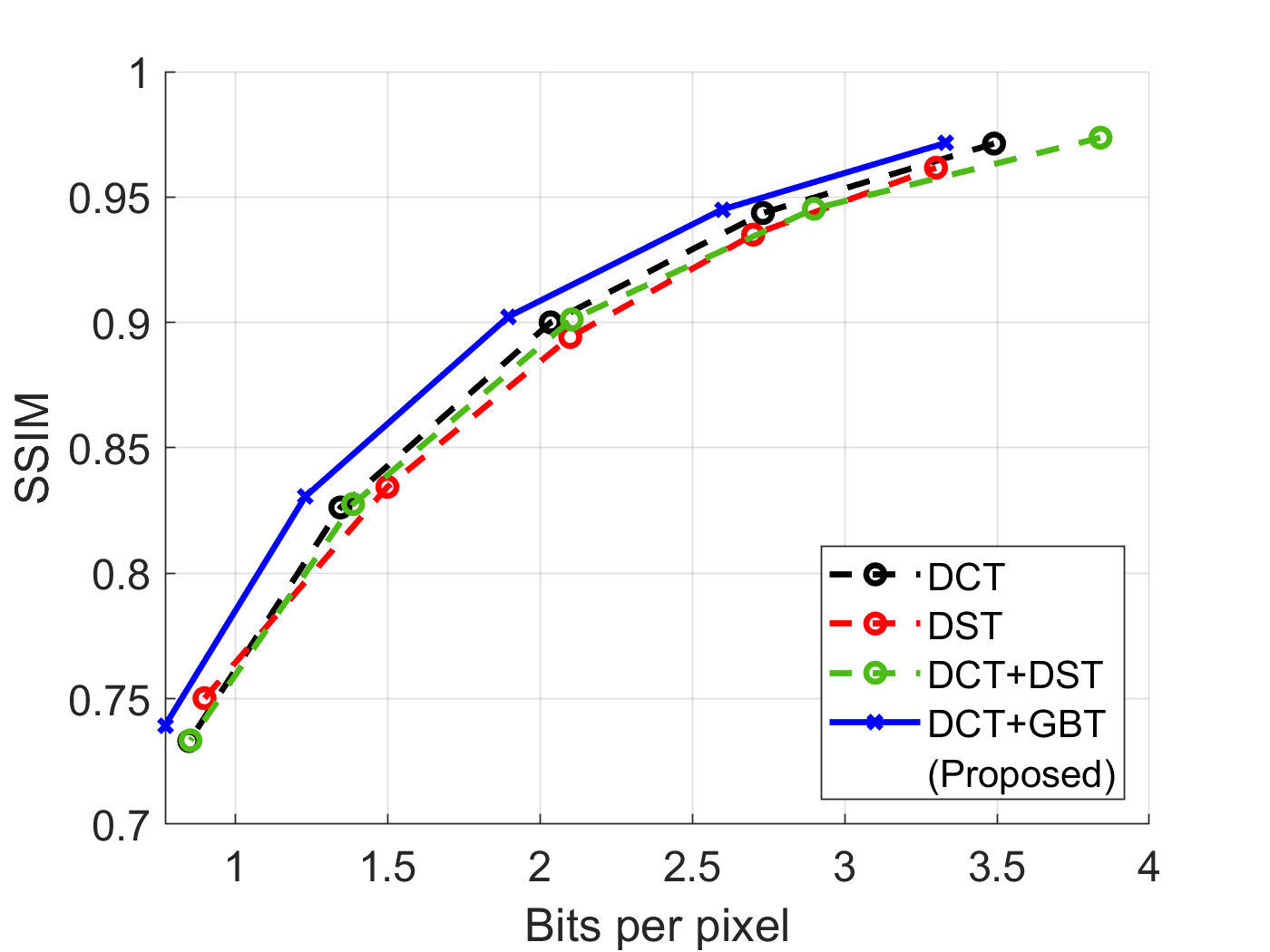}}
    \caption{Comparison of the rate-distortion curves between the proposed scenario (DCT+GBT) and other transforms along with standard H.264/AVC codec. (a) PSNR metric. (b) SSIM metric.}
    \label{fig:RDrate}
\end{figure}


The dataset is a collection of KTH\cite{caputo2005class}, Kylberg\cite{kylberg2011kylberg}, and UIUC\cite{lazebnik2005sparse} texture images, 
with 64 semantic categories and 4335 images in total. 
We collect 20 images from each category to test the performance, and all images are cropped to $320 \times 320$. 
The proposed separable path GBT is an alternative transform for the standard H.264/AVC intra-prediction, with QP values from the set ${23,27,31,35,39}$. 
The encoding block size is $16 \times 16$, and the $K$-means clustering is set at $K=8$. The GBT kernels are mode-independent and do not affect the selection of the intra-prediction modes. The residual for each block is transformed by either DCT or GBT according to the rate-distortion optimization.

\begin{figure}[t!]
    \vspace{-5mm}
    \centering
    \subfloat[\label{sfig:usagePercentage}]
        {\includegraphics[width=0.5\linewidth]{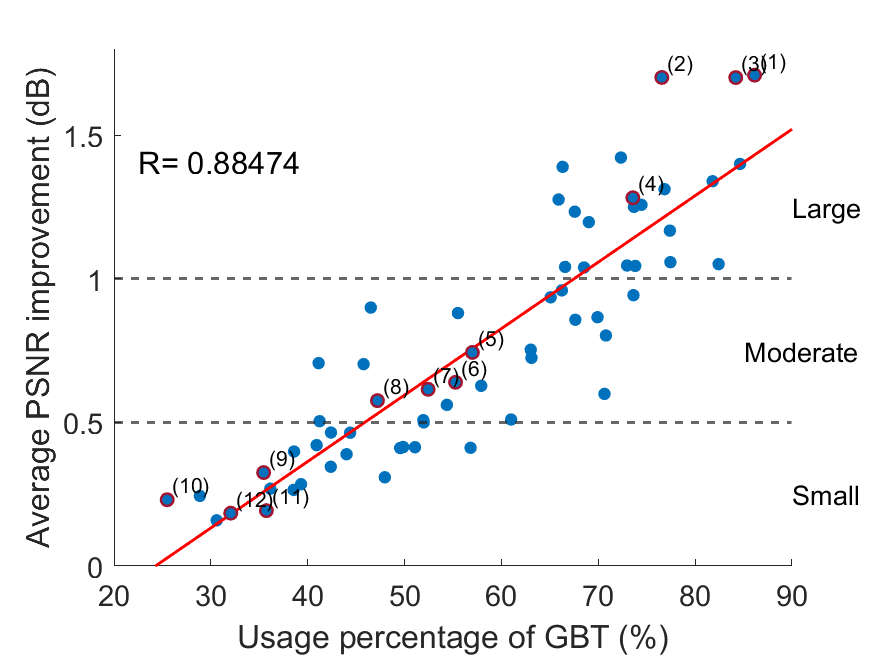}}  
    \subfloat[\label{sfig:GLNU}]
        {\includegraphics[width=0.5\linewidth]{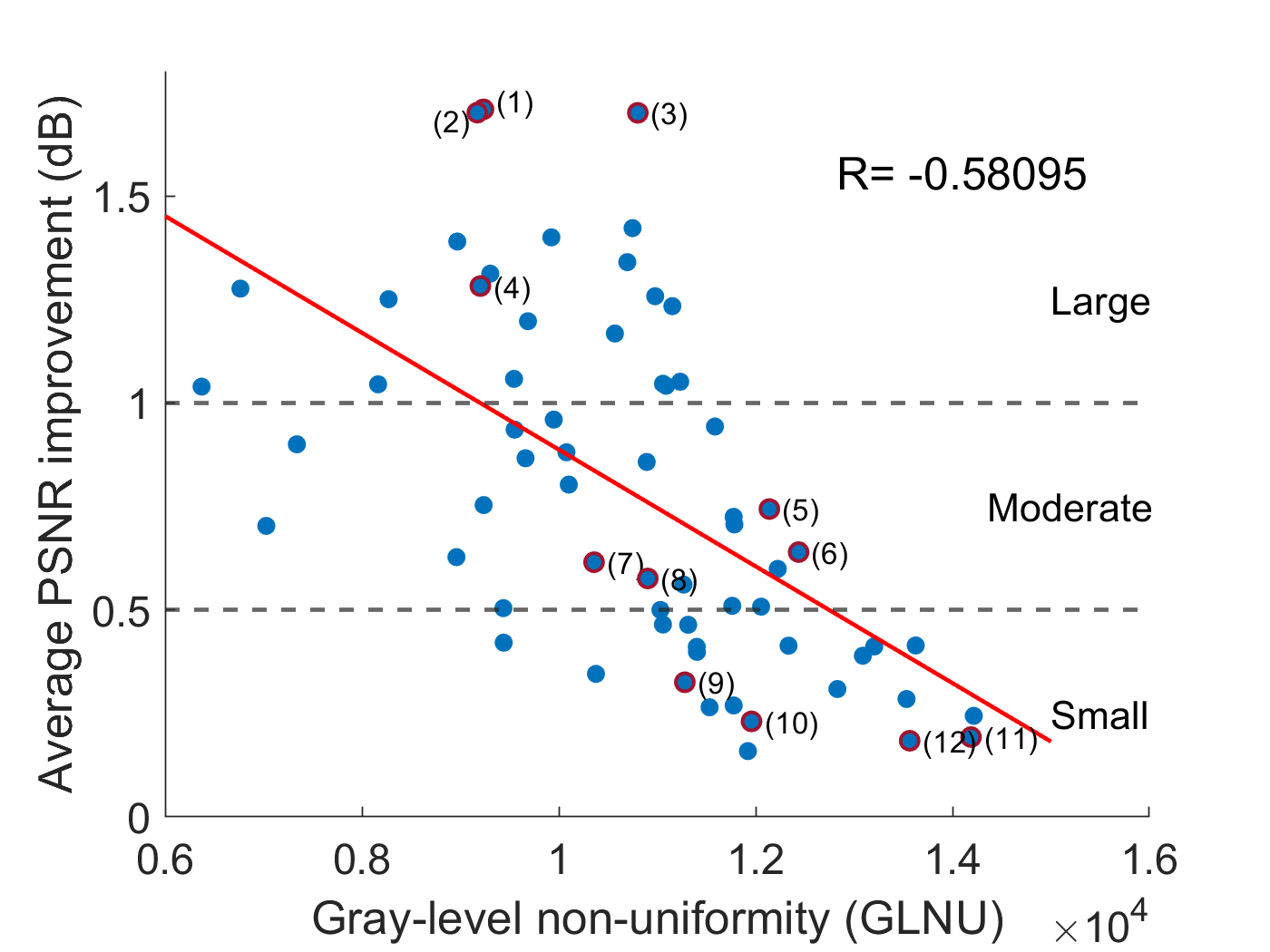}}
    \\ \vspace{-2.0mm}
    \subfloat[\label{sfig:texturesGroupByGain}]
        {\includegraphics[width=0.9\linewidth]{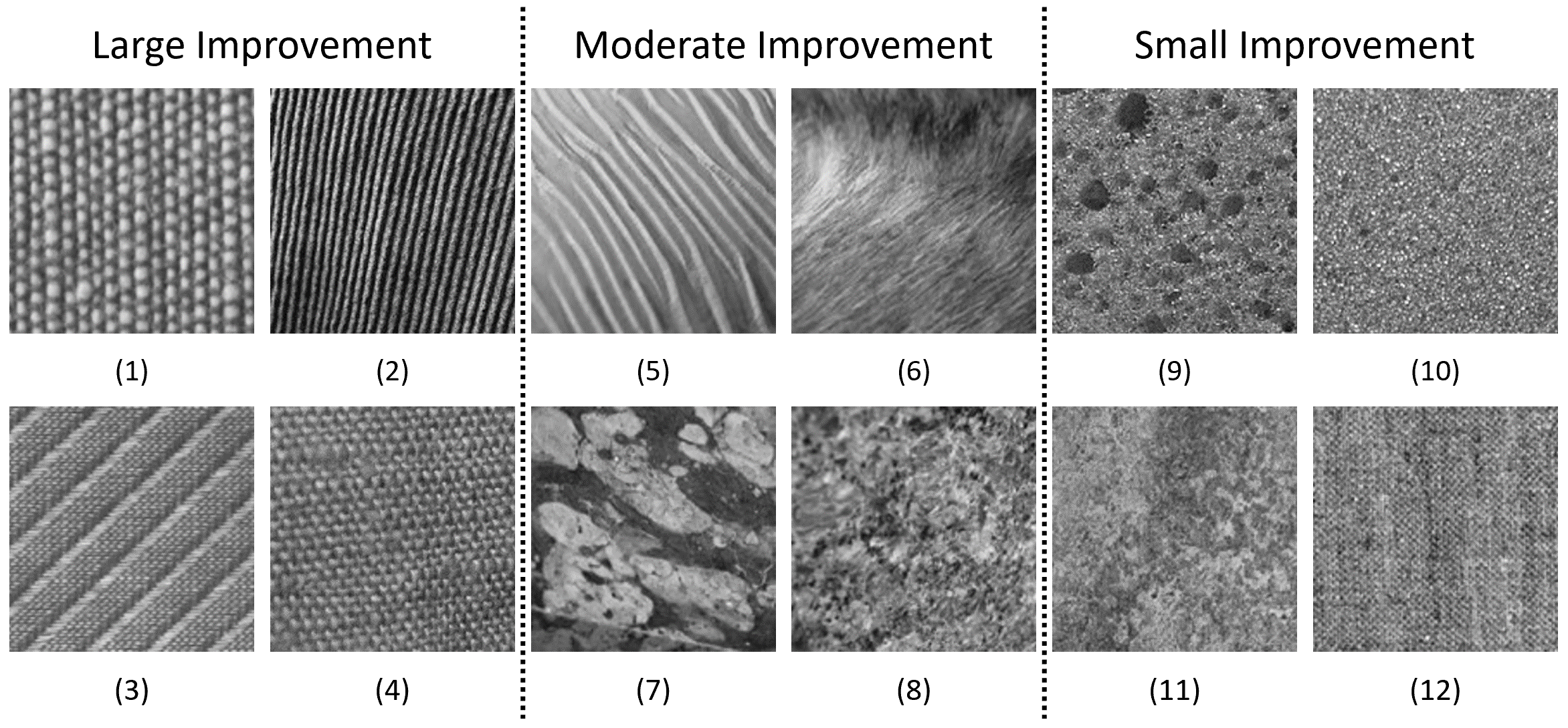}}
    \caption{Analysis on the PSNR gain from replacing the H.264/AVC DCT with the proposed scenario. Red line shows the linear relationship between the variables in x-axis and y-axis with correlation coefficient $R$. (a) Usage percentage of GBT to average PSNR improvement. (b) Gray-level non-uniformity (GLNU) to average PSNR improvement. (c) Examples of the textures grouped by PSNR improvement.}
    \label{fig:GainAnalysis}
    \vspace{-2.75mm}
\end{figure}

The proposed scenario (DCT+GBT) is compared with other data-independent transforms, including DCT and DST, along with the H.264/AVC coding standard. 
We also build a scenario (DCT+DST) that selects a better transform between DCT and DST for each block.
As can be seen in Fig.~\ref{fig:RDrate}, the proposed scenario provides better scores for PSNR and SSIM metrics using fewer bits to encode the block.
The Bjøntegaard delta bitrate saving (BD-rate) \cite{bjontegaard2001calculation} for the proposed method w.r.t the DCT is -7.8\%, which means 7.8\% bits can be saved on average compared to the standard H.264/AVC codec. 
The average PSNR improvement w.r.t the usage percentage of GBT averaged on each category is displayed in Fig.~\ref{fig:GainAnalysis}\subref{sfig:usagePercentage}, which shows a high correlation and the effectiveness of the GBT. 

To explore which types of textures are more favorable for the GBT, the gray-level non-uniformity (GLNU) derived from the gray-level run-length matrix (eq. (3) in \cite{chu1990use}) is measured and averaged on each category of texture. 
Higher GLNU values suggest a greater texture complexity or heterogeneity.
The relation between the average PSNR improvement and the GLNU is displayed in Fig.~\ref{fig:GainAnalysis}\subref{sfig:GLNU}, which shows a negative correlation suggesting that a more regular or predictable texture pattern leads to a larger gain improvement by the GBT. 
The results in Fig.~\ref{fig:GainAnalysis}\subref{sfig:usagePercentage} and Fig.~\ref{fig:GainAnalysis}\subref{sfig:GLNU} are divided into three groups based on the PSNR improvement, with some examples of textures in each group shown in Fig.~\ref{fig:GainAnalysis}\subref{sfig:texturesGroupByGain}. 
It is observed that the regular or predictable texture patterns tend to have a great gain improvement by GBT, while the stochastic or noisy texture patterns tend to have a slight improvement.

Additionally, a comparison of BD-rate savings w.r.t DCT over different image uniformity between the two SDT scenarios (DCT+GBT) and (DCT+KLT \cite{lan2011exploiting}) is drawn. 
The median GLNU value of all images categorizes the entire dataset into uniform and non-uniform images.
The results are summarized in Table~\ref{table_BDRate}. 
Although the KLT as an optimal SDT works better in general, the GBT can outperform KLT in uniform images.
This is because there are few dominant variations in uniform images, making the KLT more sensitive to visual artifacts, such as blur and ringing. By contrast, the GBT constrained to be a path graph has fewer parameters, which acts as some form of regularization, reducing visual artifacts' influence. 

\begin{table} [!t]
    \begin{center}
    \caption{Comparison of BD-Rate Savings over Different Image Uniformity}
    \label{table_BDRate}
        \begin{tabular}{| c | c | c | c |}
            \hline
              & Uniform & Non-uniform & All images \\
            \hline
            DCT+KLT\cite{lan2011exploiting} & -7.30\% & -12.56\% & -10.30\% \\
            \hline 
            \makecell{DCT+GBT  \\ (Proposed)} & -11.02\% & -4.70\% & -7.8\% \\
            \hline
        \end{tabular}
    \end{center}
    \vspace{-1.5em}
\end{table}


\section{Conclusions}
This paper introduces a novel adaptive online learning scenario for separable path-GBTs for transform coding, demonstrating improved coding efficiency for textures with complicated structures. 
The proposed scenario has a much lower computational cost than the KLT derived from template matching because the sequential $K$-means clustering reduces the number of comparisons, and the path graph edge weights are learned from data in a closed form. 
Our results show that the proposed GBT outperforms the KLT when the training samples are insufficient and has a higher coding efficiency than the DCT when the image blocks correspond to more predictable regular textures.

The proposed GBT can be used as an effective SDT combined, especially when the directional prediction fails to capture the complicated structures. 
Future work includes exploring alternative tree graph structures other than path graphs to enhance decorrelation. 
The GBT usage determined by some metrics, such as graph smoothness, before the prediction of each block, is also being considered.

\bibliographystyle{IEEEtran}
\bibliography{Ref/refs}

\begin{thebibliography}{10}
\providecommand{\url}[1]{#1}
\csname url@samestyle\endcsname
\providecommand{\newblock}{\relax}
\providecommand{\bibinfo}[2]{#2}
\providecommand{\BIBentrySTDinterwordspacing}{\spaceskip=0pt\relax}
\providecommand{\BIBentryALTinterwordstretchfactor}{4}
\providecommand{\BIBentryALTinterwordspacing}{\spaceskip=\fontdimen2\font plus
\BIBentryALTinterwordstretchfactor\fontdimen3\font minus \fontdimen4\font\relax}
\providecommand{\BIBforeignlanguage}[2]{{%
\expandafter\ifx\csname l@#1\endcsname\relax
\typeout{** WARNING: IEEEtran.bst: No hyphenation pattern has been}%
\typeout{** loaded for the language `#1'. Using the pattern for}%
\typeout{** the default language instead.}%
\else
\language=\csname l@#1\endcsname
\fi
#2}}
\providecommand{\BIBdecl}{\relax}
\BIBdecl

\bibitem{wiegand2003overview}
T.~Wiegand, G.~J. Sullivan, G.~Bjontegaard, and A.~Luthra, ``Overview of the h. 264/avc video coding standard,'' \emph{IEEE Transactions on circuits and systems for video technology}, vol.~13, no.~7, pp. 560--576, 2003.

\bibitem{ahmed1974discrete}
N.~Ahmed, T.~Natarajan, and K.~R. Rao, ``Discrete cosine transform,'' \emph{IEEE transactions on Computers}, vol. 100, no.~1, pp. 90--93, 1974.

\bibitem{pavez2022laplacian}
E.~Pavez, ``Laplacian constrained precision matrix estimation: Existence and high dimensional consistency,'' in \emph{International Conference on Artificial Intelligence and Statistics}.\hskip 1em plus 0.5em minus 0.4em\relax PMLR, 2022, pp. 9711--9722.

\bibitem{egilmez2017graph}
H.~E. Egilmez, E.~Pavez, and A.~Ortega, ``Graph learning from data under laplacian and structural constraints,'' \emph{IEEE Journal of Selected Topics in Signal Processing}, vol.~11, no.~6, pp. 825--841, 2017.

\bibitem{sullivan2012overview}
G.~J. Sullivan, J.-R. Ohm, W.-J. Han, and T.~Wiegand, ``Overview of the high efficiency video coding (hevc) standard,'' \emph{IEEE Transactions on circuits and systems for video technology}, vol.~22, no.~12, pp. 1649--1668, 2012.

\bibitem{strang1999discrete}
G.~Strang, ``The discrete cosine transform,'' \emph{SIAM review}, vol.~41, no.~1, pp. 135--147, 1999.

\bibitem{bross2021overview}
B.~Bross, Y.-K. Wang, Y.~Ye, S.~Liu, J.~Chen, G.~J. Sullivan, and J.-R. Ohm, ``Overview of the versatile video coding (vvc) standard and its applications,'' \emph{IEEE Transactions on Circuits and Systems for Video Technology}, vol.~31, no.~10, pp. 3736--3764, 2021.

\bibitem{shuman2013emerging}
D.~I. Shuman, S.~K. Narang, P.~Frossard, A.~Ortega, and P.~Vandergheynst, ``The emerging field of signal processing on graphs: Extending high-dimensional data analysis to networks and other irregular domains,'' \emph{IEEE signal processing magazine}, vol.~30, no.~3, pp. 83--98, 2013.

\bibitem{ortega2018graph}
A.~Ortega, P.~Frossard, J.~Kova{\v{c}}evi${\acute{\text{c}}}$, J.~M. Moura, and P.~Vandergheynst, ``Graph signal processing: Overview, challenges, and applications,'' \emph{Proceedings of the IEEE}, vol. 106, no.~5, pp. 808--828, 2018.

\bibitem{lan2011exploiting}
C.~Lan, J.~Xu, G.~Shi, and F.~Wu, ``Exploiting non-local correlation via signal-dependent transform (sdt),'' \emph{IEEE Journal of Selected Topics in Signal Processing}, vol.~5, no.~7, pp. 1298--1308, 2011.

\bibitem{goyal2000transform}
V.~K. Goyal, J.~Zhuang, and M.~Veiterli, ``Transform coding with backward adaptive updates,'' \emph{IEEE Transactions on Information Theory}, vol.~46, no.~4, pp. 1623--1633, 2000.

\bibitem{effros2004suboptimality}
M.~Effros, H.~Feng, and K.~Zeger, ``Suboptimality of the karhunen-loeve transform for transform coding,'' \emph{IEEE Transactions on Information Theory}, vol.~50, no.~8, pp. 1605--1619, 2004.

\bibitem{jana2006optimality}
S.~Jana and P.~Moulin, ``Optimality of klt for high-rate transform coding of gaussian vector-scale mixtures: Application to reconstruction, estimation, and classification,'' \emph{IEEE Transactions on Information Theory}, vol.~52, no.~9, pp. 4049--4067, 2006.

\bibitem{ye2008improved}
Y.~Ye and M.~Karczewicz, ``Improved h. 264 intra coding based on bi-directional intra prediction, directional transform, and adaptive coefficient scanning,'' in \emph{2008 15th IEEE International Conference on Image Processing}.\hskip 1em plus 0.5em minus 0.4em\relax IEEE, 2008, pp. 2116--2119.

\bibitem{takamura2013intra}
S.~Takamura and A.~Shimizu, ``On intra coding using mode dependent 2d-klt,'' in \emph{2013 Picture Coding Symposium (PCS)}.\hskip 1em plus 0.5em minus 0.4em\relax IEEE, 2013, pp. 137--140.

\bibitem{liu2018scene}
Y.~Liu and J.~Ostermann, ``Scene-based klt for intra coding in hevc,'' in \emph{2018 Picture Coding Symposium (PCS)}.\hskip 1em plus 0.5em minus 0.4em\relax IEEE, 2018, pp. 6--10.

\bibitem{dapena2002hybrid}
A.~Dapena and S.~Ahalt, ``A hybrid dct-svd image-coding algorithm,'' \emph{IEEE transactions on Circuits and Systems for video technology}, vol.~12, no.~2, pp. 114--121, 2002.

\bibitem{gu2011low}
Z.~Gu, W.~Lin, B.-s. Lee, and C.~Lau, ``Low-complexity video coding based on two-dimensional singular value decomposition,'' \emph{IEEE transactions on image processing}, vol.~21, no.~2, pp. 674--687, 2011.

\bibitem{cao2014singular}
X.~Cao and Y.~He, ``Singular vector decomposition based adaptive transform for motion compensation residuals,'' in \emph{2014 IEEE International Conference on Image Processing (ICIP)}.\hskip 1em plus 0.5em minus 0.4em\relax IEEE, 2014, pp. 4127--4131.

\bibitem{zhang2017signal}
T.~Zhang, H.~Chen, M.-T. Sun, D.~Zhao, and W.~Gao, ``Signal dependent transform based on svd for hevc intracoding,'' \emph{IEEE Transactions on Multimedia}, vol.~19, no.~11, pp. 2404--2414, 2017.

\bibitem{lan2017variable}
C.~Lan, J.~Xu, W.~Zeng, G.~Shi, and F.~Wu, ``Variable block-sized signal-dependent transform for video coding,'' \emph{IEEE Transactions on Circuits and Systems for Video Technology}, vol.~28, no.~8, pp. 1920--1933, 2017.

\bibitem{pavez2018learning}
E.~Pavez, H.~E. Egilmez, and A.~Ortega, ``Learning graphs with monotone topology properties and multiple connected components,'' \emph{IEEE Transactions on Signal Processing}, vol.~66, no.~9, pp. 2399--2413, 2018.

\bibitem{mateos2019connecting}
G.~Mateos, S.~Segarra, A.~G. Marques, and A.~Ribeiro, ``Connecting the dots: Identifying network structure via graph signal processing,'' \emph{IEEE Signal Processing Magazine}, vol.~36, no.~3, pp. 16--43, 2019.

\bibitem{dong2019learning}
X.~Dong, D.~Thanou, M.~Rabbat, and P.~Frossard, ``Learning graphs from data: A signal representation perspective,'' \emph{IEEE Signal Processing Magazine}, vol.~36, no.~3, pp. 44--63, 2019.

\bibitem{lu2018learning}
K.-S. Lu, E.~Pavez, and A.~Ortega, ``On learning laplacians of tree structured graphs,'' in \emph{2018 IEEE Data Science Workshop (DSW)}.\hskip 1em plus 0.5em minus 0.4em\relax IEEE, 2018, pp. 205--209.

\bibitem{egilmez2020graph}
H.~E. Egilmez, Y.-H. Chao, and A.~Ortega, ``Graph-based transforms for video coding,'' \emph{IEEE Transactions on Image Processing}, vol.~29, pp. 9330--9344, 2020.

\bibitem{lake2010discovering}
B.~Lake and J.~Tenenbaum, ``Discovering structure by learning sparse graphs,'' \emph{in Proceedings of the 33rd Annual Cognitive Science Conference}, 2010.

\bibitem{caputo2005class}
B.~Caputo, E.~Hayman, and P.~Mallikarjuna, ``Class-specific material categorisation,'' in \emph{Tenth IEEE International Conference on Computer Vision (ICCV'05) Volume 1}, vol.~2.\hskip 1em plus 0.5em minus 0.4em\relax IEEE, 2005, pp. 1597--1604.

\bibitem{kylberg2011kylberg}
G.~Kylberg, \emph{Kylberg texture dataset v. 1.0}.\hskip 1em plus 0.5em minus 0.4em\relax Centre for Image Analysis, Swedish University of Agricultural Sciences and~…, 2011.

\bibitem{lazebnik2005sparse}
S.~Lazebnik, C.~Schmid, and J.~Ponce, ``A sparse texture representation using local affine regions,'' \emph{IEEE transactions on pattern analysis and machine intelligence}, vol.~27, no.~8, pp. 1265--1278, 2005.

\bibitem{zhao2019vvc}
X.~Zhao, S.-H. Kim, Y.~Zhao, H.~E. Egilmez, M.~Koo, S.~Liu, J.~Lainema, and M.~Karczewicz, ``Transform coding in the vvc standard,'' \emph{IEEE Transactions on Circuits and Systems for Video Technology}, vol.~31, no.~10, pp. 3878--3890, 2021.

\bibitem{king2012online}
A.~King, ``Online k-means clustering of nonstationary data,'' \emph{Prediction Project Report}, pp. 1--9, 2012.

\bibitem{bjontegaard2001calculation}
G.~Bjontegaard, ``Calculation of average psnr differences between rd-curves,'' \emph{ITU SG16 Doc. VCEG-M33}, 2001.

\bibitem{chu1990use}
A.~Chu, C.~M. Sehgal, and J.~F. Greenleaf, ``Use of gray value distribution of run lengths for texture analysis,'' \emph{Pattern recognition letters}, vol.~11, no.~6, pp. 415--419, 1990.

\end{thebibliography}


\begin{thebibliography}{10}
\providecommand{\url}[1]{#1}
\csname url@samestyle\endcsname
\providecommand{\newblock}{\relax}
\providecommand{\bibinfo}[2]{#2}
\providecommand{\BIBentrySTDinterwordspacing}{\spaceskip=0pt\relax}
\providecommand{\BIBentryALTinterwordstretchfactor}{4}
\providecommand{\BIBentryALTinterwordspacing}{\spaceskip=\fontdimen2\font plus
\BIBentryALTinterwordstretchfactor\fontdimen3\font minus \fontdimen4\font\relax}
\providecommand{\BIBforeignlanguage}[2]{{%
\expandafter\ifx\csname l@#1\endcsname\relax
\typeout{** WARNING: IEEEtran.bst: No hyphenation pattern has been}%
\typeout{** loaded for the language `#1'. Using the pattern for}%
\typeout{** the default language instead.}%
\else
\language=\csname l@#1\endcsname
\fi
#2}}
\providecommand{\BIBdecl}{\relax}
\BIBdecl

\bibitem{wiegand2003overview}
T.~Wiegand, G.~J. Sullivan, G.~Bjontegaard, and A.~Luthra, ``Overview of the h. 264/avc video coding standard,'' \emph{IEEE Transactions on circuits and systems for video technology}, vol.~13, no.~7, pp. 560--576, 2003.

\bibitem{sullivan2012overview}
G.~J. Sullivan, J.-R. Ohm, W.-J. Han, and T.~Wiegand, ``Overview of the high efficiency video coding (hevc) standard,'' \emph{IEEE Transactions on circuits and systems for video technology}, vol.~22, no.~12, pp. 1649--1668, 2012.

\bibitem{bross2021overview}
B.~Bross, Y.-K. Wang, Y.~Ye, S.~Liu, J.~Chen, G.~J. Sullivan, and J.-R. Ohm, ``Overview of the versatile video coding (vvc) standard and its applications,'' \emph{IEEE Transactions on Circuits and Systems for Video Technology}, vol.~31, no.~10, pp. 3736--3764, 2021.

\bibitem{strang1999discrete}
G.~Strang, ``The discrete cosine transform,'' \emph{SIAM review}, vol.~41, no.~1, pp. 135--147, 1999.

\bibitem{zhao2019vvc}
X.~Zhao, S.-H. Kim, Y.~Zhao, H.~E. Egilmez, M.~Koo, S.~Liu, J.~Lainema, and M.~Karczewicz, ``Transform coding in the vvc standard,'' \emph{IEEE Transactions on Circuits and Systems for Video Technology}, vol.~31, no.~10, pp. 3878--3890, 2021.

\bibitem{ye2008improved}
Y.~Ye and M.~Karczewicz, ``Improved h. 264 intra coding based on bi-directional intra prediction, directional transform, and adaptive coefficient scanning,'' in \emph{2008 15th IEEE International Conference on Image Processing}.\hskip 1em plus 0.5em minus 0.4em\relax IEEE, 2008, pp. 2116--2119.

\bibitem{takamura2013intra}
S.~Takamura and A.~Shimizu, ``On intra coding using mode dependent 2d-klt,'' in \emph{2013 Picture Coding Symposium (PCS)}.\hskip 1em plus 0.5em minus 0.4em\relax IEEE, 2013, pp. 137--140.

\bibitem{liu2018scene}
Y.~Liu and J.~Ostermann, ``Scene-based klt for intra coding in hevc,'' in \emph{2018 Picture Coding Symposium (PCS)}.\hskip 1em plus 0.5em minus 0.4em\relax IEEE, 2018, pp. 6--10.

\bibitem{lan2011exploiting}
C.~Lan, J.~Xu, G.~Shi, and F.~Wu, ``Exploiting non-local correlation via signal-dependent transform (sdt),'' \emph{IEEE Journal of Selected Topics in Signal Processing}, vol.~5, no.~7, pp. 1298--1308, 2011.

\bibitem{shuman2013emerging}
D.~I. Shuman, S.~K. Narang, P.~Frossard, A.~Ortega, and P.~Vandergheynst, ``The emerging field of signal processing on graphs: Extending high-dimensional data analysis to networks and other irregular domains,'' \emph{IEEE signal processing magazine}, vol.~30, no.~3, pp. 83--98, 2013.

\bibitem{ortega2018graph}
A.~Ortega, P.~Frossard, J.~Kova{\v{c}}evi${\acute{\text{c}}}$, J.~M. Moura, and P.~Vandergheynst, ``Graph signal processing: Overview, challenges, and applications,'' \emph{Proceedings of the IEEE}, vol. 106, no.~5, pp. 808--828, 2018.

\bibitem{egilmez2017graph}
H.~E. Egilmez, E.~Pavez, and A.~Ortega, ``Graph learning from data under laplacian and structural constraints,'' \emph{IEEE Journal of Selected Topics in Signal Processing}, vol.~11, no.~6, pp. 825--841, 2017.

\bibitem{pavez2018learning}
E.~Pavez, H.~E. Egilmez, and A.~Ortega, ``Learning graphs with monotone topology properties and multiple connected components,'' \emph{IEEE Transactions on Signal Processing}, vol.~66, no.~9, pp. 2399--2413, 2018.

\bibitem{lu2018learning}
K.-S. Lu, E.~Pavez, and A.~Ortega, ``On learning laplacians of tree structured graphs,'' in \emph{2018 IEEE Data Science Workshop (DSW)}.\hskip 1em plus 0.5em minus 0.4em\relax IEEE, 2018, pp. 205--209.

\bibitem{mateos2019connecting}
G.~Mateos, S.~Segarra, A.~G. Marques, and A.~Ribeiro, ``Connecting the dots: Identifying network structure via graph signal processing,'' \emph{IEEE Signal Processing Magazine}, vol.~36, no.~3, pp. 16--43, 2019.

\bibitem{dong2019learning}
X.~Dong, D.~Thanou, M.~Rabbat, and P.~Frossard, ``Learning graphs from data: A signal representation perspective,'' \emph{IEEE Signal Processing Magazine}, vol.~36, no.~3, pp. 44--63, 2019.

\bibitem{egilmez2020graph}
H.~E. Egilmez, Y.-H. Chao, and A.~Ortega, ``Graph-based transforms for video coding,'' \emph{IEEE Transactions on Image Processing}, vol.~29, pp. 9330--9344, 2020.

\bibitem{pavez2022laplacian}
E.~Pavez, ``Laplacian constrained precision matrix estimation: Existence and high dimensional consistency,'' in \emph{International Conference on Artificial Intelligence and Statistics}.\hskip 1em plus 0.5em minus 0.4em\relax PMLR, 2022, pp. 9711--9722.

\bibitem{king2012online}
A.~King, ``Online k-means clustering of nonstationary data,'' \emph{Prediction Project Report}, pp. 1--9, 2012.

\bibitem{caputo2005class}
B.~Caputo, E.~Hayman, and P.~Mallikarjuna, ``Class-specific material categorisation,'' in \emph{Tenth IEEE International Conference on Computer Vision (ICCV'05) Volume 1}, vol.~2.\hskip 1em plus 0.5em minus 0.4em\relax IEEE, 2005, pp. 1597--1604.

\bibitem{kylberg2011kylberg}
G.~Kylberg, \emph{Kylberg texture dataset v. 1.0}.\hskip 1em plus 0.5em minus 0.4em\relax Centre for Image Analysis, Swedish University of Agricultural Sciences and~…, 2011.

\bibitem{lazebnik2005sparse}
S.~Lazebnik, C.~Schmid, and J.~Ponce, ``A sparse texture representation using local affine regions,'' \emph{IEEE transactions on pattern analysis and machine intelligence}, vol.~27, no.~8, pp. 1265--1278, 2005.

\bibitem{bjontegaard2001calculation}
G.~Bjontegaard, ``Calculation of average psnr differences between rd-curves,'' \emph{ITU SG16 Doc. VCEG-M33}, 2001.

\bibitem{chu1990use}
A.~Chu, C.~M. Sehgal, and J.~F. Greenleaf, ``Use of gray value distribution of run lengths for texture analysis,'' \emph{Pattern recognition letters}, vol.~11, no.~6, pp. 415--419, 1990.

\end{thebibliography}

\end{document}